\centerline  {\bf Repairing the algebraic foundations of the Standard Model of particle physics}
\vskip5pt
\centerline  {Douglas Newman}
\vskip5pt
\centerline {e-mail: \it dougnewman276@gmail.com}
\vskip20pt

\beginsection Abstract

The Standard Model (SM) of particle physics is in such good 
agreement with experiment that it is still accepted as providing
an accurate model of reality, with the role of chiral symmetry in electro-weak 
unification regarded as one of its major achievements. Nevertheless, its
conceptual and algebraic foundation is faulty.  Chirality
is shown to be algebraically inconsistent with neutrinos having
finite mass and being fermions. 

\vskip50pt

\beginsection \S1. Introduction 

In spite of the widely acclaimed successes of the Standard Model (SM), 
important questions are raised by its conceptual and algebraic foundations. 
These were constructed over a considerable period time and, as new experimental
results became available, successive additions
led to its present very complicated form. This has resisted attempts 
to incorporate it into a unified theory that leaves its algebraic foundations intact. 
The usual justification for retaining the SM in the analysis of 
high energy experimental results is that neutrino masses
make a negligible contribution to relativistic momentum/energy
conservation. Direct tests of the SM 
are centred on experimental searches for deviations from
its numerical predictions, leaving open the possibility that experimental 
results can be misinterpreted when they are analysed using a faulty algebraic model.
 
Faults in the algebraic foundations of the SM are:

\item {1.} It retains its initial assumption that neutrinos have
zero mass.

\item {2.} It introduces the concept of chirality which distinguishes
two types of neutrino (denoted by L and R), when
only L neutrinos have been observed.

\item {3.} It fails to predict the existence of different
neutrino generations.   

Previous work has raised doubts about the mathematical formulation
of the SM. Stoica [3] has shown that the theoretical reasons 
for introducing chiral symmetry breaking are far from obvious,
but does not find any reason to reject it.  
Dartora and Cabrera [4] showed that much of electro-weak theory
can be explained in terms of the $Cl_{3,3}$ algebra if
chirality is omitted. Recent work by Gording and Schmidt-May 
[5] follows Furey [6] in assigning fermions
to the elements of an algebra isomorphic to the complex Clifford
algebra $Cl_6$. Their approach, which (apart from employing a 
Clifford algebra) produces an algebraic basis for the SM, but
does not justify the introduction of chirality.

The necessary repairs to the algebraic foundations
of the SM are carried out as follows:

\item{\S2} analyses the original reason for 
introducing chiral symmetry breaking into the SM, 
and shows it to ensure that neutrinos cannot be fermions.

\item {\S3} summarises the consequences of this analysis.
\vskip5pt

The Appendix clarifies the physical interpretation of Clifford algebras
and their matrix representations in Clifford Unification, which replaces
the SM. 

\beginsection \S2. Chirality

The SM was initially (and remains) based on the assumption that neutrinos
are zero mass fermions that move at the velocity of light. This produces
in-consistences with the standard description of fermions in terms of 
quantum numbers that are defined in a coordinate system in which they are at rest
(e.g.[7]\S4.6.1).

Specifically, the SM replaces the time ($\gamma^0$) coordinate, that
distinguishes fermions from anti-fermions in the Dirac theory, with chirality ($\gamma^5$)
that distinguishes between observed L neutrinos, denoted L, with 
eigenvalue $\mu_5= -1$, and unobserved R neutrinos denoted R, with eigenvalue $\mu_5= +1$.
Such a physical distinction is made possible because $\gamma^5$ is invariant
under Lorentz transformations (as it commutes with all the $\gamma^{\mu\nu}$)
making the division into L and R parts of Dirac 4-spinors independent of 
the space-time coordinate frame. This is currently incorporated into SM matrix element
calculations by including the projection operator $P_L= {1\over 2}(1+\gamma^5)$ in
matrix elements. The algebraic distinction between neutrinos and anti-neutrinos is lost,
losing contact with experimental results that distinguish neutrinos anti-neutrinos 
(e.g see [2]\S8).

$\gamma^5$ commutes with the elements $\gamma^{0\mu}$ of $Cl_{1,3}$ for $\mu = 1,2,3$ 
that describe the direction of motion (in the three coordinate directions) of particle
moving at the velocity of light. Motion in arbitrary directions is described by linear
combinations of the  $\gamma^{0\mu}$. The problem with this description is that the 
spin orientation of a neutrino moving at the velocity of light is necessarily parallel to its 
direction of motion. For example, a neutrino moving at the velocity of light in the
direction $\gamma^1$ is described by $\gamma^{01}$. This only commutes with the 
spin orientation $\gamma^{23}$. This is not a possible description of fermions
which have spin orientations independent of their direction of motion.
(It is worth noting that this provides a perfectly good description of photons: a photon
moving in the direction $\gamma_1$ has velocity $\gamma^{01}$, which 
commutes with $\gamma^{23}$, but not with $\gamma^{12}$ and $\gamma^{13}$, 
relating its direction of motion with two possible planes of polarization.)  
 
Another problem with $\gamma^5$ is that it does not commute with
$\gamma^0$, conflicting with Dirac interpretation of $\gamma^0$ as
the operator that distinguishes fermions from anti-fermions. This ensures
that neutrinos cannot be described by usual form of Dirac 4-spinors that distinguish fermions 
in terms of quantum numbers determined by the eigenvalues of fermion rest frame coordinates.
Nevertheless, it is possible to express chirality as a quantum number with modified
forms of 4-spinor as given, for example, by [7]\S6.4. These forms replace $\gamma^0$
with $\gamma^5$, eliminating the normal means of distinguishing fermions and anti-fermions.
An important feature of the SM is that the weak field only acts on the L-chiral 
part of {\it all} fermions, which form the SU(2) doublets listed at the top
of page 416 of [7]. Their non-interacting R-chiral parts are SU(2) singlets.

The resulting confusion in getting a Lie group description of neutrinos continues 
to produce considerable amount of theoretical discussion as to whether they
should be described by Weyl, Majorana or Dirac wave functions (e.g. [8] \S1.4.1).
A recent example is [9], which claims to show that neutrinos should be described 
by Majorana wave functions. However, this calculation is invalidated because it is
based on the SM. Many analyses of experimental data involve anti-neutrinos,
which are not defined in the SM (e.g. see [2]\S8).

An argument for retaining the SM description of neutrinos as 
zero mass chiral particles has been that their mass is so small that 
it makes negligible difference in analyses of high energy 
experiments that are based on the conservation of energy-momentum 4-vectors.
This is supported theoretically by the close relationship between 
4-spinors with chirality and helicity quantum numbers, e.g. see [7]\S6.4.2.

The transition from SM to CU descriptions of processes that involve the
the projection operator $P_L= {1\over 2}(1+\gamma^5)$ are made by simply
omitting it from the formulae in which it occurs. This should not affect
the results of many calculations, giving a possible reason why the SM has
been so successful.

\beginsection\S3. Summary 

This analysis of the algebraic foundations
of the Standard Model has shown that:\vskip2pt

\item{\bf 1.} Its assumption that observed neutrinos 
are all L-chiral (with eigenvalues -1 of $\gamma^5$) is 
inconsistent with the fact that neutrinos have
finite mass and require the normal 4-spinor description of fermions
based on the Dirac equation. The predicted R-chiral neutrinos 
have never been observed.

\item{\bf 2.} Chirality is the main reason that it has not been found
possible to incorporate the SM in unified field theories. 

\item{\bf 3.} Omission of the projection operator 
$P_L= {1\over 2}(1+\gamma^5)$ in many calculations
should not affect their results, possibly explaining the success of the SM.

\item{\bf 4.} The arguments leading to the V-A potential as a description
parity non-conservation in weak interactions are invalid for two reasons: 
(1) it depends on the chiral description of the weak interaction and 
(2) it is based on the incorrect assumption that all fermions have the same
intrinsic parity [1,2]. 

\item{\bf 5.} Attempts to retain the algebraic 
and conceptual structures of the SM in a unified theory 
are bound to fail: a complete revision is necessary, 
such as that provided by Clifford Unification in [2].

\vskip50pt

\beginsection Appendix: The physical interpretation of the Clifford algebras

Clifford algebras $Cl_{p,q}$ have p generators $\gamma^k$ where $(\gamma^k)^2 = 1$
and q generators $\gamma^j$ where $(\gamma^j)^2 = -1$. The generators all anti-commute.
Thiw work is concerned with $Cl_{1,3}$ in which 
the single `p' generator $\gamma^0$ interpreted as the unit time interval,
and the three `q' generators
$\gamma^1, \gamma^2, \gamma^3$ interpreted as unit displacements
in the three orthogonal directions of space. Its applications
in quantum mechanics are currently based on the identification of commuting
elements the algebra with different physical properties of fermions. The original 
example of this is the $Cl_{1,3}$ Dirac algebra where 
the eigenvalues of $\gamma^0$ distinguish
between fermions and anti-fermions. Some linear combination of 
$\gamma^{\mu\nu},(\mu,\nu = 1,2,\>{\rm or}\> 3)$, which all commute
with $\gamma^0$, define the spin orientation and have eigenvalues
that determine the spin direction. 

Matrix representations of all the $Cl_{p,q}$ can be constructed.
For example $Cl_{1,3}$ has 4$\times$4 matrix representations, 
including the representation constructed by Dirac. The actual choice of
these matrices is restricted only in that they satisfy the 
relations between elements of the algebra. Otherwise they have
no inherent physical significance, but the choice of matrix 
representation does determine the physical interpretation of 
elements of the spinors in the Dirac equation. 

Representation matrices of the elements of Clifford algebras can 
be expressed as Kronecker products of the three real $2\times 2$ matrices
$$ 
{\bf P}  =\left(\matrix{ 0 & -1\cr 1 & 0\cr} \right)= -i\sigma_2,\>
\>{\bf Q}=\left(\matrix{ 0 & 1\cr 1 & 0\cr} \right) = \sigma_1,\>
\>{\bf R}=\left(\matrix{ -1 & 0\cr 0 & 1\cr} \right) = -\sigma_3 .      \eqno(A.1)                        
$$
 $\bf P,Q,R$ satisfy 
$$
-{\bf P}^2 = {\bf Q}^2 = {\bf R}^2 = {\bf I}, \>\>
{\bf P}{\bf Q} = {\bf R} = -{\bf Q}{\bf P},\>{\bf P}{\bf R} 
= -{\bf	Q} = -{\bf R}{\bf P}, \>{\bf Q}{\bf R} = -{\bf P} 
= -{\bf R}{\bf Q}.                                                      \eqno (A.2)
$$
Practical advantages of this notation, rather than the familiar 
Pauli $\sigma$ matrices, are the avoidance of suffices and the provision of a clear
distinction between real and complex matrices.

The 4$\times$4 Dirac matrices
employed in the SM (e.g. see Chapter 4 of [7]) correspond to the Kronecker products
$$
\gamma^0= -{\bf I}\otimes {\bf R},\>\gamma^1= -{\bf Q}\otimes {\bf P},\>
\gamma^2= -i{\bf P}\otimes {\bf P},\>\gamma^3= {\bf R}\otimes {\bf P},
\>\gamma^5= -{\bf I}\otimes {\bf Q},                                      \eqno (A.3)
$$
where $\gamma^\mu (\mu = 0,1,2,3)$ are the generators of $Cl_{1,3}$.
$-i\gamma^5$ is the four dimensional unit volume. $\gamma^5$ is invariant
under Lorentz transformations.
In Dirac's theory the $\gamma^\mu$ define coordinate systems
that are at rest with respect to a specific fermion. The
quantum  mechanical interpretation of Dirac's algebra is based on describing 
fermions with the commuting elements of $Cl_{1,3}$, with eigenvalues
that distinguish the four different states of the same fermion 
(e.g. see Chapter 4 of [7]). The Dirac equation, expressed
in terms of 4-component spinors, has four solutions 
with eigenvalues $\pm1$ of $\gamma^0$ that
distinguish electrons from positrons or, more generally,
fermions from anti-fermions. Arbitrary spatial orientations
of spin defined by linear combinations of the $\gamma^{\mu\nu}, {\mu,\nu=1,2,3}$,
with eigenvalues defining its two possible directions. 
For example, the two eigenvalues of $\gamma^{12}$ describe 
opposite directions of a spin with $\gamma^3$ orientation.

Fermion properties are distinguished by two diagonal matrices in
the Dirac algebra. In the above representation $\gamma^0 =-{\bf I}\otimes {\bf R}$ 
corresponds to time intervals (e.g.\S4.6.1 of [7]); 
spin (orientated in the $\gamma^3$ direction) is
represented by $\gamma^{12} = i{\bf R}\otimes {\bf I}$ (e.g. \S4.4 of [7]).

Both $Cl_{1,3}$ is a subalgebra
of $Cl_{3,3}$ which, as shown in [2], provides a complete description
of the relationship between fermion properties and space-time,
defining three of the seven quantum numbers that distinguish 
all the elementary fermions in Clifford unification (CU).  
The relationship between the quantum mechanical fermions and
their space-time coordinates is always expressed in
terms of a `rest frame' fixed on the fermion. A specific choice of this is the
matrix representation of $Cl_{3,3}$ listed in Appendix B of [2].

\vskip80pt 
\beginsection References

\frenchspacing

\item {[1]} Newman, Douglas (2021) Unified theory of elementary fermions and their interactions
based on Clifford algebras {arXiv:2108.08274}

\item {[2]} Newman, D.J. (2024) Quantum number conservation: a tool in the design and analysis
of high energy experiments J.Phys.G:Nucl.Part.Phys. 51 095002\vskip1pt
Corrigendum: J.Phys.G:Nucl.Part.Phys. 52,(2025) 019501 

\item{[3]} Stoica, Ovidiu Cristinel. (2020) Chiral asymmetry in the weak interaction via Clifford Algebras \vskip 1pt {arXiv:2005.08855}

\item{[4]} Dartora, C.A. and Cabrera, G.G. (2009) The Dirac equation and a non-chiral electroweak theory
in six dimensional space-time from a locally gauged $SO(3,3)$ symmetry group {arXiv: 0901.4230v1} \vskip 1pt
Int. J. Theor. Phys. (2010) {\bf 49}:51-61  

\item{[5]} Gording, Brage and Schmidt-May, Agnis. (2020) The Unified Standard Model. 
{arXiv:1909.05641} 

\item{[6]} Furey, C. (2018) Three generations, two unbroken gauge symmetries, and one eight-dimensional algebra. Phys. Lett. B {\bf 785} 84 

\item {[7]} Thomson, Mark. (2013) Modern Particle Physics (Cambridge University Press)

\item {[8]} Ather, M. Sajjad and Singh, S.K. (2020) The Physics of Neutrino Interactions. (Cambridge University Press)

\item {[9]} Bhatt, Vivan; Mondal, Rajrupa; Vaibhav, Vatsalya and Singh, Tejinder P. (2023) \vskip 1pt
Majorana Neutrinos, Exceptional Jordan Algebra and Mass Ratios for Charged Fermions \vskip 1pt 
J.Phys.G::Nucl. Part. Phys. {\bf 49} 045007 (2022) {arXiv:2108.05787} 

\item {[11]} Bettini, Alessandro. (2008) Introduction to Elementary Particle Physics
(Cambridge University Press)

\item {[12]} Cottingham, W.N. and Greenwood, D.A. (2008) An Introduction \vskip1pt
to the Standard Model of Particle Physics, Second Edition. (Cambridge University Press)

\item {[13]} Aitchison, I. J. R. and Hey, A. J. G.(2004)  Gauge Theories in 
Particle Physics, Volume II: QCD and the Electroweak Theory (IOP Publishing Ltd)

\end

\bye